\documentclass[twocolumn,preprintnumbers,prl,showpacs,superscriptaddress,prl]{revtex4-1}

\usepackage{palatino}
\usepackage[latin1]{inputenc}
\usepackage{epsf}
\usepackage{amsmath,amssymb}
\usepackage{latexsym}
\usepackage{calc}
\usepackage{color}
\usepackage{graphicx}
\usepackage{float}

\DeclareMathAlphabet\mathbfcal{OMS}{cmsy}{b}{n}

\def\bA{{\bf A}}
\def\bP{{\bf P}}

\def\dulR{{\underline{\underline{\bf R}}}}
\def\dulr{{\underline{\underline{\bf r}}}}

\begin{document}

\title{A coupled-trajectory quantum-classical approach to 
{electronic} decoherence in non-adiabatic processes}
\author{Seung Kyu Min}
\affiliation{Max-Planck Institut f\"ur Mikrostrukturphysik, Weinberg 2,
D-06120 Halle, Germany}
\affiliation{Department of Chemistry, School of Natural Science, Ulsan 
National Institute of Science and Technology (UNIST), Ulsan 689-798, Korea}
\author{Federica Agostini}
\affiliation{Max-Planck Institut f\"ur Mikrostrukturphysik, Weinberg 2,
D-06120 Halle, Germany}
\author{E. K. U. Gross}
\affiliation{Max-Planck Institut f\"ur Mikrostrukturphysik, Weinberg 2,
D-06120 Halle, Germany}

\date{\today}
\begin{abstract}
We present a novel quantum-classical approach to non-adiabatic dynamics, deduced from the coupled electronic and nuclear equations in the framework of the exact factorization of the electron-nuclear wave function. The method is based on the quasi-classical interpretation of the nuclear wave function, whose phase is related to the classical momentum and whose density is represented in terms of classical trajectories. In this approximation, electronic decoherence is naturally induced as effect of the coupling to the nuclei and correctly reproduces the expected quantum behaviour. Moreover, the splitting of the nuclear wave packet is captured as consequence of the correct approximation of the time-dependent potential of the theory. This new approach offers a clear improvement over Ehrenfest-like dynamics. The theoretical derivation presented in the Letter is supported by numerical results that are compared to quantum mechanical calculations.
\end{abstract}
\maketitle

The theoretical description of phenomena such as vision~\cite{cerulloN2010, *schultenBJ2009, *ishidaJPCB2012}, photo-synthesis~\cite{tapaviczaPCCP2011, *flemingN2005}, photo-voltaic processes~\cite{rozziNC2013, *silvaNM2013, *jailaubekovNM2013}, proton-transfer and hydrogen storage~\cite{sobolewski, *varella, *hammes-schiffer, *marx} is among the most challenging problems in Condensed Matter Physics and Theoretical Chemistry. The underlying quantum dynamics of electrons and nuclei exhibit a non-adiabatic character, meaning that it cannot be explained by employing the Born-Oppenheimer (BO) approximation. In this respect, the major challenge for theory resides in the explicit treatment of electronic excited-state dynamics coupled to the nuclear motion. While methods that retain quantum features of the nuclear dynamics~\cite{martinezCPL1996, *martinezJPC1996, *martinezACR2006, *cederbaumCPL1990, *burghardtJCP1999, *worthTCA2003, *thossCM2004, *thossJCP2003, *thossJPCC2010, *mengJCP2012, *brownPCCP} are the most accurate to address this problem, they cannot be applied to systems with hundreds, or even thousands, of atoms. Therefore, a treatment of nuclear dynamics in terms of (semi)classical trajectories~\cite{ehrenfest, *shalashilin_JCP2009, *kapral-ciccotti, *ivano, *bonellaJCP2005, *marxPRL2002, *pechukas, *subotnik, *herman, *coker, *prezhdo, *Thoss_PRL1997, *Miller_JCP1997_2, *truhlarFD2004, *ciccottiJCP2000, *kapralJCP2013, *cokerJCP2012, *martinezACR2006, *wyattJCP2001, *burghardtJCP2005, *prezhdoPRL2001, *tannorJCP2012, *burghardtJCP2011, *thossJCP2013, *ananthJCP2013, *io, mqc, long_mqc, tully1990} represent the most promising and numerically feasible approach for actual calculations. Despite the great effort that has been devoted over the years to the development of such methods, actual applications are still limited~\cite{subotnikJCP2011, *landryJCP2011, *subotnikJCP2011_1, *subotnikJCP2011_2,  *tavernelliJCP2013, *worthJMGM2007, *robbJPCA2003, *tully1998}. Well-known issues are connected to the lack of, or incorrect account for, decoherence and to the inability of reproducing the spatial splitting of a nuclear wave packet, as in Ehrenfest-like dynamics. In the study of electronic non-adiabatic processes, these problems can result in wrong predictions for quantum populations and in unphysical outcomes for the nuclear dynamics. 

We have recently proposed a new formalism that can be employed to overcome the above issues, the so-called exact factorization of the electron-nuclear wave function~\cite{AMG, *AMG2}. In this framework, the full wave function is written as the product of a nuclear wave function and an electronic factor with a parametric dependence on the nuclear configuration. Coupled equations drive the dynamics of the two components of the wave function. In particular, a time-dependent Schr\"odinger equation (TDSE) describes the evolution of the nuclear wave function where the effect of the electrons, beyond BO, is accounted for in a \textit{single}, time-dependent, potential. Compared to a formulation in terms of multiple static adiabatic (or BO) potential energy surfaces (PESs), the advantage of this formulation is evident: when the classical approximation is introduced, the force driving the nuclear evolution can be \textit{uniquely} determined from the gradient of this time-dependent potential.~\cite{footnote_potentials}

In previous work we have: (i) analyzed the features of the time-dependent potential~\cite{steps} in the context of non-adiabatic proton-coupled-electron-transfer, in order to pinpoint the properties that need to be accounted for when introducing approximations; (ii) determined the suitability of the classical and quasi-classical treatment~\cite{long_steps, *long_steps_mt} of nuclear dynamics, in a situation where the electronic effect can be taken into account exactly; (iii) derived an independent-trajectory (IT) mixed quantum-classical (MQC) algorithm~\cite{mqc, long_mqc} to solve the coupled electronic and nuclear equations (from the factorization) in a fully approximate way. In particular, the IT-MQC scheme has been obtained as the \textsl{lowest-order} approximation, in an expansion in powers of $\hbar$ of the nuclear wave function in the complex-phase representation. Further investigation~\cite{semiclassics} has shown, however, that corrections are required if the nuclei exhibit a quantum behavior related to a non-adiabatic event, e.g. the splitting of a nuclear wave packet after the passage through an avoided crossing.

The aim of this Letter is to go beyond the IT-MQC algorithm of Ref.~\cite{mqc, long_mqc}. We have derived a coupled-trajectory (CT) MQC algorithm able to reproduce the features of the time-dependent potential, by evolving an ensemble of classical trajectories to mimic the quantum evolution of the nuclei. Electronic populations, decoherence and spatial splitting of the nuclear wave packet are correctly reproduced when the new scheme is employed, as will be demonstrated below.

The exact factorization approach consists in writing the solution, $\Psi(\dulr,\dulR,t)$, of the TDSE $\hat H\Psi=i\hbar\partial_t\Psi$, as the single product $\Psi(\dulr,\dulR,t)=\Phi_\dulR(\dulr,t)\chi(\dulR,t)$, where $\Phi_\dulR(\dulr,t)$ is an electronic factor parametrically depending on the nuclear positions and $\chi(\dulR,t)$ is a nuclear wave function. Here, $\hat H=\hat T_n+\hat H_{BO}$ is the Hamiltonian describing the system of interacting electrons and nuclei, with $\hat T_n$ the nuclear kinetic energy and $\hat H_{BO}$ the BO Hamiltonian containing all interactions among the particles and the electronic kinetic energy. The positions of $N_e$ electrons and $N_n$ nuclei are represented by the symbols $\dulr$ and $\dulR$, respectively. The product-form of $\Psi$ is unique, up to within an $(\dulR,t)$-dependent gauge-like transformation, if the partial normalization condition, $\int d\dulr|\Phi_{\dulR}(\dulr,t)|^2=1$ $\forall\,\dulR,t$, is imposed. The evolution of the two components of the full wave function is governed by an electronic equation, 
\begin{align}\label{eqn: electronic equation}
\left[\hat H_{BO}+\hat U_{en}[\Phi_\dulR,\chi]-\epsilon(\dulR,t)\right]\Phi_{\dulR}=i\hbar\partial_t\Phi_{\dulR},
\end{align}
and a nuclear equation, 
\begin{align}\label{eqn: nuclear equation}
\left[\sum_{\nu=1}^{N_n}\frac{[-i\hbar\nabla_\nu+\bA_\nu(\dulR,t)]^2}{2M_\nu}
+\epsilon(\dulR,t)\right]\chi=i\hbar\partial_t\chi,
\end{align}
which are exactly equivalent to the full TDSE. The electron-nuclear coupling operator,
\begin{align}
\label{eqn:encoupling}
&\hat U_{en}[\Phi_\dulR,\chi]=\sum_{\nu=1}^{N_n}\frac{[-i\hbar\nabla_\nu{-}\bA_{\nu}(\dulR,t)]^2}{2M_\nu}\nonumber\\ 
&+\frac{1}{M_\nu}\left(\frac{-i\hbar\nabla_\nu\chi}{\chi}+\bA_{\nu}(\dulR,t)\right)\Big(-i\hbar\nabla_\nu-\bA_{\nu}(\dulR,t)\Big),
\end{align}  
represents the effect of the nuclei on electronic dynamics; in turn, the time-dependent vector potential, 
\begin{align}\label{eqn: vector potenttial}
\bA_{\nu}(\dulR,t)=\langle\Phi_\dulR(t)|-i\hbar\nabla_\nu\Phi_\dulR(t)\rangle_\dulr,
\end{align}
and time-dependent PES (TDPES),
\begin{align}
\label{eqn:exactTDPES}
\epsilon(\dulR,t)=\langle\Phi_\dulR(t)|\hat H_{BO}+\hat U_{en}-i\hbar\partial_t|\Phi_\dulR(t)\rangle_\dulr, 
\end{align}
account for the electronic back-reaction on the nuclei in a Schr\"odinger-like equation. These potentials are uniquely determined~\cite{AMG, *AMG2} up to within a gauge transformation.

The CT-MQC scheme adopts a description of nuclear dynamics in terms of classical trajectories, $\dulR^{(I)}(t)$, thus all quantities depending on $\dulR,t$ will become functions of $\dulR^{(I)}(t),t$. Nuclear dynamics will be \textsl{sampled} using trajectories, meaning that we track the evolution of a nuclear wave packet by looking at how the trajectories evolve in time. Information about the nuclear space $\dulR$ is only available at the instantaneous positions along the classical paths. It follows that we will not be able to calculate partial time derivatives, but only total time derivatives, by using the chain rule $d/dt=\partial_t+\sum_{\nu}\mathbf V_\nu^{(I)}\cdot\nabla_\nu$, with $\mathbf V_\nu^{(I)}=\dot{\mathbf R}_\nu^{(I)}(t)$ the nuclear velocity. Henceforth, the superscript $(I)$ will be used to indicate a spatial dependence, e.g., $\bA_\nu^{(I)}(t)=\bA_\nu(\dulR^{(I)}(t),t)$.

The main steps in the derivation of the new CT-MQC scheme are the following: (a) we approximate the TDPES, to avoid expensive calculations of second-order derivatives of the electronic wave function with-respect-to the nuclear coordinates; (b) we fix the gauge freedom; (c) we introduce a quasi-classical interpretation of the nuclear wave function, whose phase is connected to the classical momentum and modulus reconstructed in terms of Gaussian wave packets; (d) we expand the electronic wave function on the adiabatic basis (Born-Huang expansion), $\Phi^{(I)}(t)=\sum_l|C_l^{(I)}(t)|\exp[(i/\hbar)\gamma^{(I)}_l(t)]\varphi_l^{(I)}$, hence a set of partial differential equations for the coefficients of the expansion will be coupled to the nuclear equation. {Also the full wave function can be expanded on the adiabatic basis, with coefficients $F_l(\dulR,t)$, for the exact expression, or $F_l^{(I)}(t)$, for the quantum-classical case, and will be referred to as BO-projected wave packets.}

(a) In the expression of the TDPES we neglect the contribution of $\langle\Phi_\dulR(t)|\hat U_{en}|\Phi_\dulR(t)\rangle_\dulr$. Notice that the expectation value on $\Phi_\dulR$ of the second line of Eq.~(\ref{eqn:encoupling}) is zero by construction, thus the neglected term in the expression of the TDPES contains the second-order variations of the electronic state with-respect-to the nuclear coordinates, which is small~\cite{handyjcp1986, valeevjcp2003, footnote_gauge} compared to the first-order. Therefore, the TDPES is approximated as $\epsilon(\dulR,t)\simeq\epsilon_0(\dulR,t)+\epsilon_{\mathrm{TD}}(\dulR,t)$ and Eqs.~(\ref{eqn: electronic equation}) and~(\ref{eqn: nuclear equation}) become
\begin{align}
\label{eqn:approx electronic equation}
i&\hbar\dot \Phi^{(I)} = \hat{H}_{BO}\Phi^{(I)}{-}\sum_{\nu=1}^{N_n}\frac{\boldsymbol{\mathcal P}_\nu^{(I)}}{M_\nu}\!\cdot\!\left(\mathbf A^{(I)}_\nu{+i\hbar}\nabla_\nu\right)\Phi^{(I)}
\end{align}
and
\begin{align}
\label{eqn:approx nuclear equation}
&\mathbf F_\nu^{(I)} = -\left\langle\Phi^{(I)}\right|\left(\nabla_\nu \hat H_{BO}\right)\left|\Phi^{(I)}\right\rangle_\dulr\nonumber\\
&+\sum_{\nu'=1}^{N_n}\frac{2i\boldsymbol{\mathcal P}_{\nu'}^{(I)}}{\hbar M_{\nu'}}\left(\mathbf A_{\nu'}^{(I)}\mathbf A_{\nu}^{(I)}-\hbar^2\Re\left\langle\nabla_{\nu'}\Phi^{(I)}\Big|\nabla_\nu\Phi^{(I)}\right\rangle_\dulr\right),
\end{align}
respectively, where the symbol $\dot \Phi^{(I)}$ is used to indicate the full time-derivative of the electronic wave function and $\boldsymbol{\mathcal P}_\nu^{(I)}$ will be specified below. The equations have been cast in such a way that the first terms on the right-hand-side are exactly the same as in the Ehrenfest scheme~\cite{tully}. The additional terms are corrections, whose effect will be now investigated.

(b) In deriving these expressions for the evolution of the electronic wave function, Eq.~(\ref{eqn:approx electronic equation}), and for the classical nuclear force, Eq.~(\ref{eqn:approx nuclear equation}), the gauge freedom has fbeen fixed by imposing $\epsilon_0^{(I)}(t)+\epsilon_{\mathrm{TD}}^{(I)}(t)+\sum_\nu\mathbf V_\nu^{(I)}\cdot\bA_\nu^{(I)}(t)=0$.

(c) The corrections beyond-Ehrenfest in Eqs.~(\ref{eqn:approx electronic equation}) and~(\ref{eqn:approx nuclear equation}) contain a term $\boldsymbol{\mathcal P}_\nu^{(I)}(t)=-i\hbar\nabla_{\nu} |\chi^{(I)}(t)|/|\chi^{(I)}(t)|$, which we will refer to as \textsl{quantum momentum}. The reason for this choice lies in the following expression
\begin{align}\label{eqn: imaginary correction to P}
\frac{-i\hbar\nabla_\nu\chi(\dulR,t)}{\chi(\dulR,t)}+\bA_\nu(\dulR,t)\simeq\mathbf P_\nu^{(I)}(t)+\boldsymbol{\mathcal P}_\nu^{(I)}(t)
\end{align}
for the term in Eq.~(\ref{eqn:encoupling}) that explicitly depends on the nuclear wave function. Such term has to be approximated when a trajectory-based treatment is adopted. In fact, Eq.~(\ref{eqn: imaginary correction to P}) has been obtained by writing the nuclear wave function in polar form, $\chi=|\chi|e^{iS/\hbar}$, and then identifying (quasi-classically) $\nabla_\nu S+\bA_\nu=\bP_\nu$, with $\bP_\nu$ the classical nuclear momentum and $\dot {\mathbf P}_\nu={\mathbf F}_\nu$ from Eq.~(\ref{eqn:approx nuclear equation}).

(d) If compared to the Ehrenfest scheme, the implementation of the CT-MQC algorithm based on Eqs.~(\ref{eqn:approx electronic equation}) and~(\ref{eqn:approx nuclear equation}) requires only two additional steps: the calculation of (i) $\nabla_\nu\Phi^{(I)}$ and (ii) $\mathcal P^{(I)}_\nu$. We employ the Born-Huang expansion of $\Phi^{(I)}$, in order to express the term (i) using the derivatives of the expansion coefficients, indicated by the symbols $C_l^{(I)}(t)$, and the non-adiabatic coupling vectors. The approximation $\nabla_\nu C_l^{(I)}\simeq (i/\hbar)\nabla_\nu\gamma_l^{(I)}C_l^{(I)}$ used here, with $\gamma_l^{(I)}(t)$ the phase of $C_l^{(I)}(t)$, is consistent with previous analysis reported in Refs.~\cite{steps, long_steps, *long_steps_mt}. Moreover, based on quasi-classical considerations described in detail in the Supplemental Material, we further approximate  $\nabla_\nu\gamma_l^{(I)}(t)\simeq-\int^t d\tau\nabla_\nu\epsilon_{BO}^{(l),(I)}$. The term (ii) is calculated assuming that the nuclear density is a combination of Gaussian-shaped wave packets, each corresponding to a given adiabatic state. Notice that this approximation is not used in general in the algorithm, but only to estimate the quantum momentum. For a two-state model, $\boldsymbol{\mathcal P}_\nu^{(I)}$ becomes~\cite{semiclassics} a linear function in the region where $\rho_l^{(I)}(t)=|C_l^{(I)}(t)|^2\neq0,1$, while it is set to zero elsewhere (see the Supplemental Material and the discussion below). The generalization of this approximation to multiple states is straightforward and will be presented elsewhere~\cite{mqc_mt}. The parameters of such linear function are the slope and the $y$-intercept, where the former is determined analytically by using Gaussian-shaped nuclear wave packets and the latter is obtained by enforcing (the reasonably physical condition) that no population exchange occurs when the non-adiabatic coupling vectors are zero. Information about the positions of all trajectories at a given time is required when evaluating these two parameters, thus resulting in a procedure beyond the IT-MQC approach: the classical trajectories cannot be evolved independently from each other, they are \textsl{coupled}.

The major advantage of the CT-MQC scheme developed here is that this procedure naturally incorporates \textit{decoherence} effects. In the following we shall discuss this feature in detail. After the nuclear wave packet has left a region of strong non-adiabatic coupling, the population $\rho_l^{(I)}(t)=|C_l^{(I)}(t)|^2$ of the $l$-th BO state changes in time as
\begin{align}
\label{eqn:rho dynamics}
\dot\rho_l^{(I)}=& -\sum_{\nu=1}^{N_n}\frac{2i\boldsymbol{\mathcal P}_\nu^{(I)}}{\hbar M_\nu}\cdot\left(\mathbf A^{(I)}_\nu-\nabla_\nu \gamma_l^{(I)}\right)\rho_l^{(I)}.
\end{align}
In this region, the expression of the vector potential reduces to $\mathbf A_\nu^{(I)}(t)=\sum_l\rho_l^{(I)}(t)\nabla_\nu\gamma_l^{(I)}(t)$, since the non-adiabatic coupling vectors are negligible. In Eq.~(\ref{eqn:rho dynamics}) we observe that, once $\rho_l^{(I)}(t)$ has approached the values 0 or 1, the term on the right-hand-side becomes zero, thus the electronic population remains constant (to 0 or 1) $\forall\,l$. This is a clear indication of \textit{decoherence}, since the (squared-modulus of the) off-diagonal elements of the electronic density matrix, often used as a measure of electronic coherence, become zero. Therefore, the correction terms beyond-Ehrenfest in Eqs.~(\ref{eqn:approx electronic equation}) and~(\ref{eqn:approx nuclear equation}), proportional to the quantum momentum, will be referred to as \textsl{decoherence terms}. Obtaining this feature is a clear improvement over the Ehrenfest approach and, likewise, over the IT-MQC approach~\cite{mqc, long_mqc} deduced from the exact factorization. Decoherence naturally appears by including dominant corrections in the expression of the nuclear wave function, leading to the appearance of the quantum momentum.

\begin{figure}
\begin{center}
  \includegraphics*[width=.4\textwidth,angle=270]{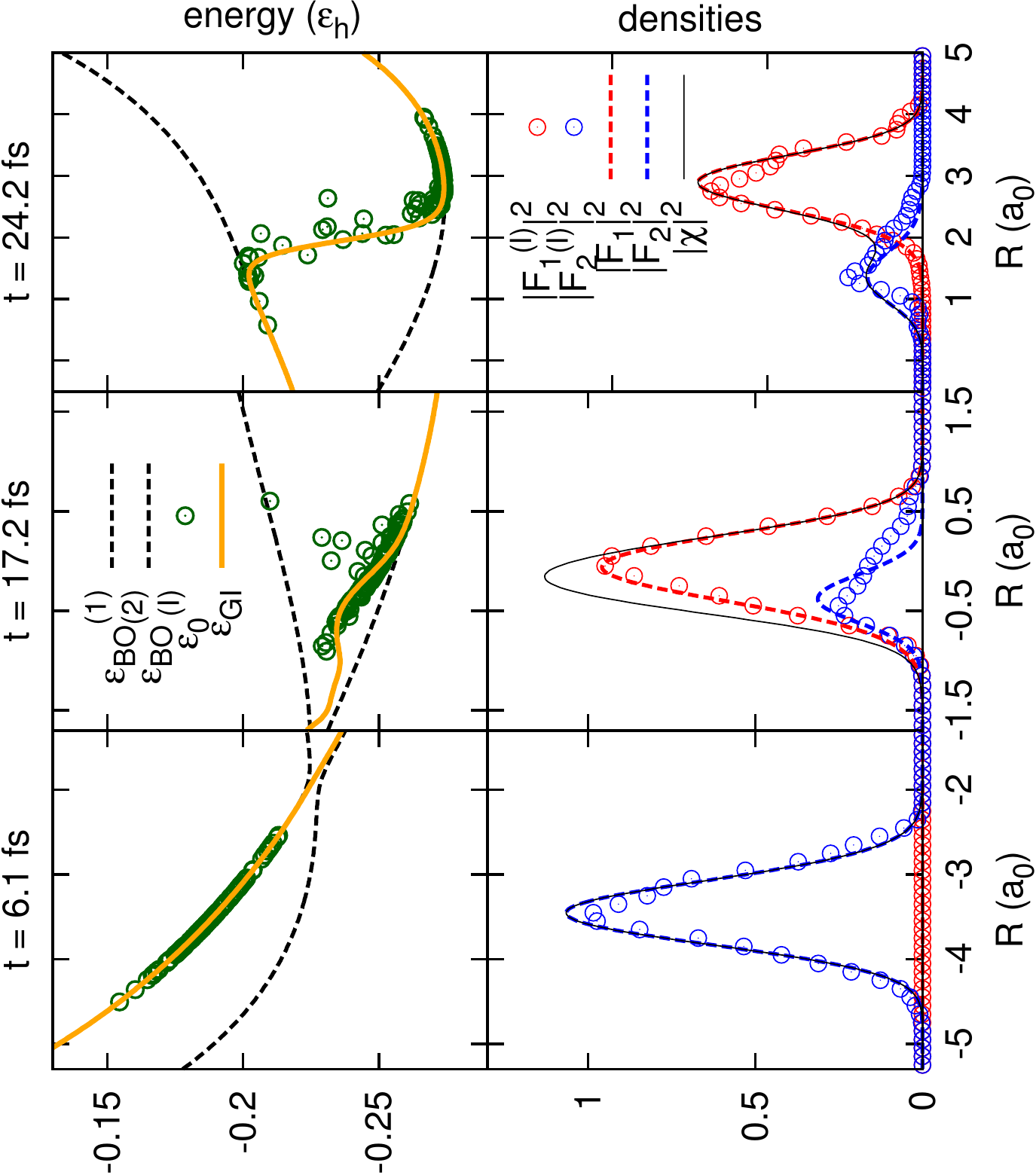}
 \end{center}
 \caption{\label{fig: model 1}Snapshots at different times for model case (1) of: (top) the gauge-invariant (GI) part of the scalar potential $\epsilon_{GI}(\dulR,t)=\langle\Phi_\dulR(t)|\hat H_{BO}+\hat U_{en}|\Phi_\dulR(t)\rangle_\dulr$ (lines) compared to its approximation $\epsilon_{0}^{(I)}(t)$ (dots) used in the CT-MQC calculations, in Hartree $\epsilon_h$ (the BO surface are plotted for reference as dashed black lines); (bottom) nuclear density $|\chi(\dulR,t)|^2$ (black line) and BO-projected densities (dashed lines) compared to the values of the BO-projected densities along the trajectories (dots).}
\end{figure}

\begin{figure}
\begin{center}
  \includegraphics*[width=.45\textwidth]{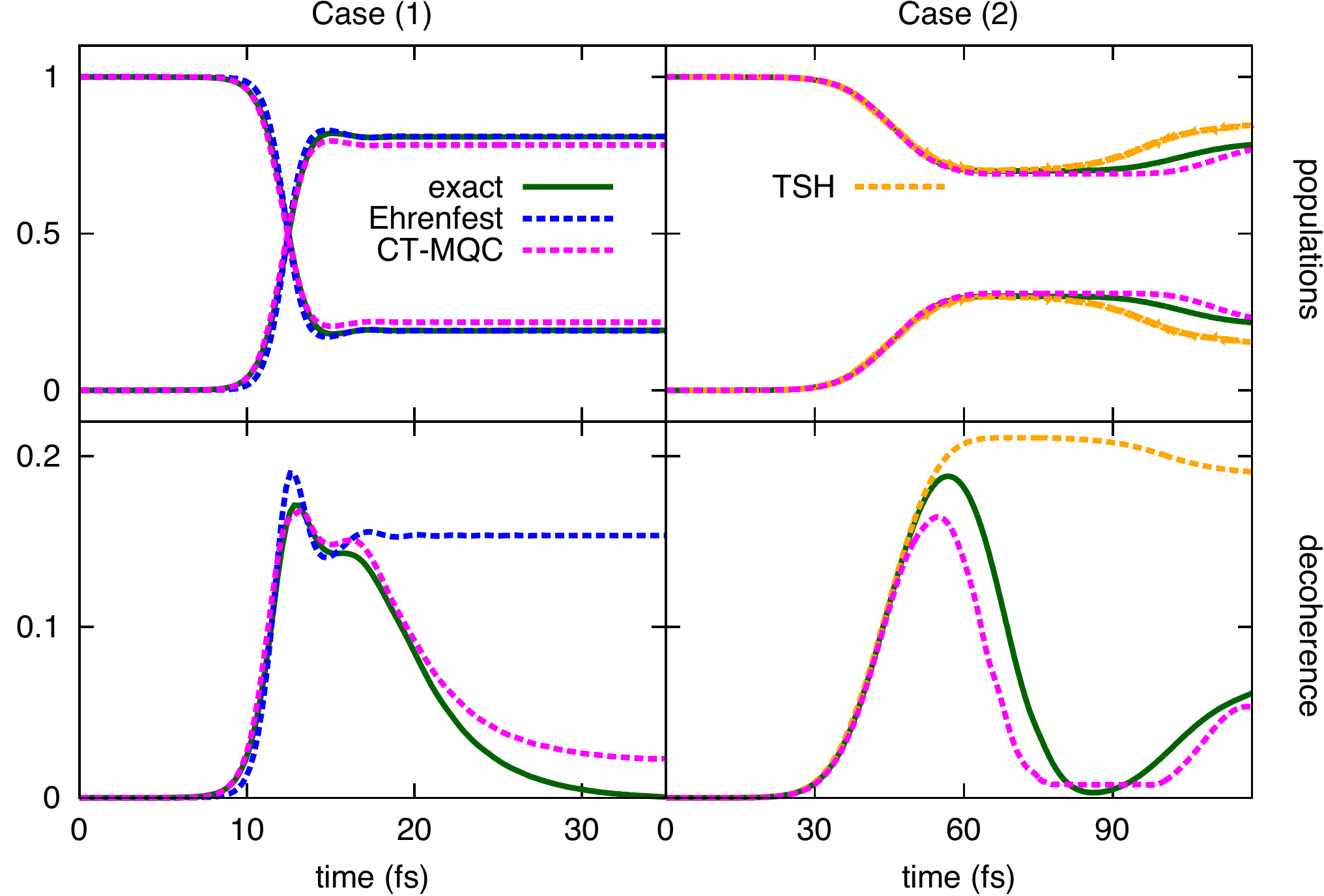}
 \end{center}
 \caption{\label{fig: populations} Populations of the BO states (upper panels) and indicator of decoherence (lower panels) as functions of time for model case (1) (left) and model case (2) (right).}
\end{figure}

Numerical results obtained by implementing the above-described method are 
shown below in comparison to exact calculations. {We discuss the 
performance of the CT-MQC algorithm in comparison to Ehrenfest dynamics for a 
two-state problem~\cite{MM} involving the passage of the nuclear wave packet 
through a single avoided crossing, case (1), and in comparison to trajectory 
surface hopping (TSH) for a two-state problem involving the reflection of the 
nuclear wave packet from a potential barrier and its consequent spatial 
splitting, case (2), commonly known as Tully-3 model~\cite{tully1990}. The computational details for these 
two problems are given in the Supplemental Material.

The TDPES for model case (1) is shown in Fig.~\ref{fig: model 1} (upper 
panels). It develops steps and the nuclear wave packet correctly splits at the 
avoided crossing (Fig.~\ref{fig: model 1}, lower panels). It is worth 
noting that Ehrenfest dynamics completely misses the splitting, as we have 
shown in Ref.~\cite{long_mqc}. Furthermore, despite the fact that Ehrenfest 
dynamics properly reproduces the populations of the electronic states, as 
shown in Fig.~\ref{fig: populations} (upper left panel), it does not capture 
decoherence. On the contrary, the CT-MQC procedure slightly underestimates the 
non-adiabatic population exchange but correctly reproduce decoherence 
(Fig.~\ref{fig: populations}, lower left panel).}

In Fig.~\ref{fig: populations}, we have used the quantity $N_{traj}^{-1}\sum_{I}\rho_1^{(I)}(t)\rho_2^{(I)}(t)$ as measure of decoherence, whose quantum equivalent is $\int d\dulR\,\rho_1(\dulR,t)\rho_2(\dulR,t)|\chi(\dulR,t)|^2$. Here, the nuclear density has been replaced by its ``classical'' approximation, i.e. $|\chi(\dulR,t)|^2\simeq N_{traj}^{-1}\sum_I\delta(\dulR-\dulR^{(I)}(t))$.

\begin{figure}

\begin{center}
 \includegraphics*[width=.45\textwidth]{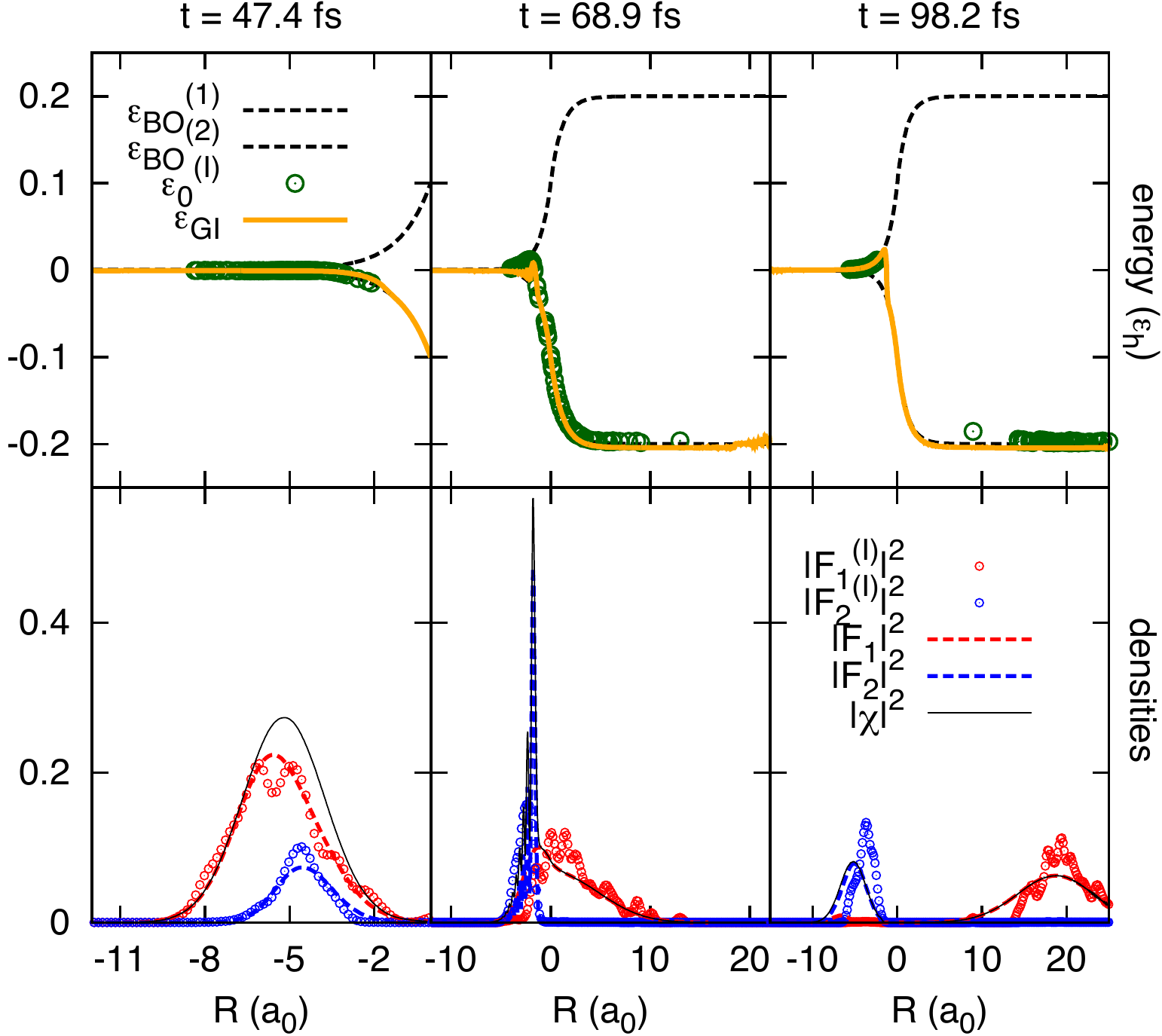}
 \end{center}
 \caption{\label{fig: model 2}Same as Fig.~\ref{fig: model 1} for model case (2).}
 \end{figure}

{We show in Fig.~\ref{fig: model 2} (lower panels) that the CT-MQC 
algorithm reproduces the splitting of the nuclear wave packet due to the 
reflection from the barrier. In fact, the TDPES develops the 
well-studied~\cite{steps} steps (Fig.~\ref{fig: model 2}, upper panels) that 
bridge piecewise adiabatic shapes and allow trajectories in different regions 
of space to feel different forces. This feature is the strength of a procedure 
based on the exact factorization: a \textit{single} time-dependent potential generating 
very different forces in different regions of space. It is 
known~\cite{tully1990} that TSH as well is able to capture the reflection 
event for a low initial momentum of the nuclear wave packet, but suffers from 
over-coherence~\cite{subotnikJCP2011_1, subotnikJCP2013}. In fact, as shown in 
Fig.~\ref{fig: populations}, TSH completely misses decoherence, whereas the 
CT-MQC scheme not only reproduces the populations of the electronic states as 
functions of time (Fig.~\ref{fig: populations}, upper right panel), but can 
also capture electronic decoherence (Fig.~\ref{fig: populations}, lower right 
panel). The comparison between exact and CT-MQC results is overall remarkable 
and a clear step forward in comparison to other methods.}


In this Letter we have proposed a CT-MQC scheme based on the exact factorization formalism and tested it on a typical example of electronic non-adiabatic process. The resulting equations give additional terms compared to Ehrefenst dynamics, that appear to be responsible for decoherence. The comparison of the CT-MQC scheme with full quantum mechanical results shows that we can correctly predict both electronic and nuclear properties: population dynamics, nuclear wave packet splitting and decoherence. Non-adiabatic transitions are induced by the classical nuclear momentum, the zero-th order term of the $\hbar$-expansion of the nuclear wave function, and decoherence is the effect of the dominant corrections to the momentum. In addition, we have proven that, as discussed in our previous work~\cite{steps, long_steps, *long_steps_mt}, being able to catch the main features of the time-dependent potential in an approximate scheme results in the correct description of the nuclear dynamics. The major advantages of our CT-MQC algorithm over commonly used methods are: (1) the working equations are conceptually and computationally as simple as Ehrenfest equations, and (2) a small number of trajectories is required, because \textit{only} initial conditions are to be sampled (no stochastic element is introduced). Working in the framework of the exact factorization allows to systematically improve previous approximations, as we have shown in this Letter in comparison to the IT-MQC of Ref.~\cite{mqc, long_mqc}. Along similar lines, future work will focus on including quantum nuclear effects, such as interference, adopting a semiclassical representation of nuclear dynamics.

Partial support from the Deutsche Forschungsgemeinschaft (SFB 762) and from the European Commission (FP7-NMP-CRONOS) is gratefully acknowledged.


\begin{thebibliography}{73}
\bibitem [{foo({\natexlab{b}})}]{footnote_gauge}%
  \BibitemOpen
  \href@noop {} {}\ \bibinfo {note} {{W}e note that
  $\Phi_\dulR(\dulr,t)$ may contain a translational phase like $e^{i{\underline
  {\underline { \mathbf K}}}\cdot\dulR}$ where ${\underline {\underline
  {\mathbf K}}}$ is the linear momentum associated to the motion of the nuclear
  system as a rigid body. The second derivative with respect to ${\mathbf
  R}_\nu$ then yields a term proportional to ${\mathbf K}_\nu^2$ which is
  proportional to $M_\nu^2$. However, the vector potential in the neglected
  term of the TDPES contains a similar term which cancels out all $M_\nu^2$
  dependencies.}\BibitemShut {Stop}%
\end{thebibliography}
%

\end{document}


\title{Supplemental Material to ``A coupled-trajectory quantum-classical approach to electronic decoherence in non-adiabatic processes''}
\author{Seung Kyu Min}
\affiliation{Max-Planck Institut f\"ur Mikrostrukturphysik, Weinberg 2,
D-06120 Halle, Germany}
\affiliation{Department of Chemistry, School of Natural Science, Ulsan 
National Institute of Science and Technology (UNIST), Ulsan 689-798, Korea
}
\author{Federica Agostini}
\affiliation{Max-Planck Institut f\"ur Mikrostrukturphysik, Weinberg 2,
D-06120 Halle, Germany}
\author{E. K. U. Gross}
\affiliation{Max-Planck Institut f\"ur Mikrostrukturphysik, Weinberg 2,
D-06120 Halle, Germany}

\date{\today}
\maketitle

In the Letter we state that the main evolution equations of the coupled-trajectory mixed quantum-classical (CT-MQC) scheme can be cast such that corrections terms to Ehrenfest-like equations arise naturally when the \textsl{quantum momentum} is included. Here, we will present this result by employing the Born-Huang expansion to represent the electronic wave function $\Phi_\dulR(\dulr,t)$ as a superposition of adiabatic states. Alternative expressions of Eqs.~(6) and~(7) of the Letter will be derived. Notice that this operation is simply a way to re-write Eqs.~(6) and~(7), no additional approximations are considered, rather than those already discussed in the Letter.

\section{Equations employing the Born-Huang expansion}
The adiabatic states are the eigenvectors of the Born-Oppenheimer (BO) Hamiltonian, $\hat H_{BO}$, and are indicted by the symbols $\lbrace\varphi_\dulR^{(l)}(\dulr)\rbrace_{l=1,N_{st}}$, where $N_{st}$ is the number of states that will be included in the expansion. The corresponding eigenvalues are labeled as $\epsilon_{BO}^{(l)}(\dulR)$. The Born-Huang expansion is
\begin{align}\label{eqn: born-huang}
\Phi_\dulR\left(\dulr,t\right)=\sum_{l=1}^{N_{st}}\left|C_l\left(\dulR,t\right)\right|\,\exp{\left[\frac{i}{\hbar}\gamma_l\left(\dulR,t\right)\right]}\varphi_\dulR^{(l)}(\dulr),
\end{align}
where the coefficients $C_l(\dulR,t)$ have been written in terms of their moduli, $|C_l(\dulR,t)|$, and phases, $\gamma_l(\dulR,t)$.

The evolution equations for the coefficients $C_l(\dulR,t)$ and the classical nuclear force, from Eqs.~(6) and~(7), can be written now as
\begin{align}
\dot C_l^{(I)}(t) =& \frac{-i}{\hbar}\epsilon_{BO}^{(l),(I)}C_l^{(I)}(t)-\sum_{\nu=1}^{N_n}\mathbf V^{(I)}(t)\cdot\sum_{k=1}^{N_{st}} C_k^{(I)}(t)\mathbf d_{\nu,lk}^{(I)} \label{eqn: el. ehrenfest}\\
&+\sum_{\nu=1}^{N_n}\frac{1}{M_\nu}\frac{\nabla_\nu |\chi^{(I)}(t)|}{|\chi^{(I)}(t)|}\cdot\left(\mathbf A_{\nu}^{(I)}(t)-\mathbf f_{\nu,l}^{(I)}(t)\right)C_l^{(I)}(t)\label{eqn: el. additional term}
\end{align}
and
\begin{align}
\mathbf F_\nu^{(I)}(t) =& -\sum_{l=1}^{N_{st}} \rho_l^{(I)}(t)\nabla_\nu\epsilon_{BO}^{(l),(I)} -\sum_{k>l,=1}^{N_{st}}2\Re\left[\rho_{lk}^{(I)}(t)\right]\left(\epsilon_{BO}^{(k),(I)}-\epsilon_{BO}^{(l),(I)}\right)\mathbf d_{\nu,lk}^{(I)}\label{eqn: n. ehrenfest} \\
&+\sum_{\nu'=1}^{N_{st}}\frac{2}{M_{\nu'}}\frac{\nabla_{\nu'} |\chi^{(I)}(t)|}{|\chi^{(I)}(t)|}\cdot \boldsymbol{\mathcal F}_{\nu'\nu}^{(I)}(t),\label{eqn: n. additional term}
\end{align}
with $\nabla_\nu|\chi^{(I)}(t)|/|\chi^{(I)}(t)|=(i/\hbar)\boldsymbol{\mathcal P}_\nu^{(I)}(t)$. Here, in order to adopt the same notation used in the main text, we indicate the $\dulR$-dependence by introducing the label $(I)$, the trajectory, and dropping the label $\dulR$. New symbols have been introduced, namely
\begin{align}
\rho_{l}^{(I)}(t) = \left|C_l^{(I)}(t)\right|^2 \quad\mbox{and}\quad\rho_{lk}^{(I)}(t) = {C_l^{(I)}}^*(t)C_l^{(I)}(t),
\end{align}
the standard non-adiabatic coupling vectors, defined as
\begin{align}
\mathbf d_{\nu,lk}^{(I)} = \left\langle \varphi^{(l),(I)}\right|\left.\nabla_\nu\varphi^{(k),(I)}\right\rangle_\dulr,
\end{align}
the gradient of the phases, $\gamma_l^{(I)}(t)$, of the expansion coefficients in Eq.~(\ref{eqn: born-huang})
\begin{align}
\mathbf f_{\nu,l}^{(I)}(t) = \nabla_\nu\gamma_l^{(I)}(t)
\end{align}
and the matrix
\begin{align}
\boldsymbol{\mathcal F}_{\nu'\nu}^{(I)}(t) = \sum_{l=1}^{N_{st}}\rho_l^{(I)}(t)\mathbf f_{\nu',l}^{(I)}(t)\wedge\left(\mathbf A_{\nu}^{(I)}(t)-\mathbf f_{\nu,l}^{(I)}(t)\right).
\end{align}
In this last expression, the symbol $\wedge$ stands for a tensor product. Moreover, notice that, using the Born-Huang expansion in Eq.~(\ref{eqn: born-huang}), also the vector potential can be (exactly) re-written as
\begin{align}
\mathbf A_{\nu}^{(I)}(t) = \sum_{l=1}^{N_{st}}\rho_l^{(I)}(t)\mathbf f_{\nu,l}^{(I)}(t)+2\hbar\sum_{k>l,=1}^{N_{st}}\Im\left[\rho_{lk}^{(I)}(I)\right]\mathbf d_{\nu,lk}^{(I)},
\end{align}
where real electronic BO states have been assumed.

Eqs.~(\ref{eqn: el. ehrenfest}) and~(\ref{eqn: n. ehrenfest}) are exactly the expressions appearing in the Ehrenfest algorithm, when the Born-Huang expansion is used to represent the electronic wave function. The additional terms depend on the \textsl{quantum momentum}, defined as
\begin{align}
\boldsymbol{\mathcal P}_\nu^{(I)}(t) = -i\hbar\frac{\nabla_{\nu} \left|\chi^{(I)}(t)\right|}{\left|\chi^{(I)}(t)\right|}
\end{align}
in Eq.~(8) of the Letter. 

Eqs.~(\ref{eqn: el. ehrenfest}) to~(\ref{eqn: n. additional term}) are the basic equations of the CT-MQC algorithm derived and discussed in the Letter.

\section{Approximation for the gradient of the expansion coefficients}
The quantity $\mathbf f_{\nu,l}^{(I)}(t)$ appears in the evolution equations as consequence of the following approximations. The gradient of the expansion coefficients in Eq.~(\ref{eqn: born-huang}), namely
\begin{align}
\nabla_\nu C_l^{(I)}(t) = \left[\frac{i}{\hbar}\nabla_\nu\gamma_{l}^{(I)}(t) + \frac{\nabla_\nu\left|C_l^{(I)}(t)\right|}{\left|C_l^{(I)}(t)\right|}\right]C_l^{(I)}(t)
\end{align}
is approximated as
\begin{align}\label{eqn: gradient of the coefficients}
\nabla_\nu C_l^{(I)}(t) \simeq \frac{i}{\hbar}C_l^{(I)}(t)\nabla_\nu\gamma_{l}^{(I)}(t),
\end{align}
since we have already observed, based on full quantum calculations~\cite{steps, long_steps, long_steps_mt}, that the contribution $\nabla_\nu\left|C_l^{(I)}(t)\right|$ is considerably different from zero only in regions where the nuclear density is small and thus where only few trajectories are expected. The error introduced this way can be systematically reduced by increasing the number of trajectories.

In the calculations, we use an approximation to the term $\nabla_\nu\gamma_{l}^{(I)}(t)$ based on the following quasi-classical considerations. When the Born-Huang expansion is also employed for the full wave function $\Psi(\dulr,\dulR,t)$, namely
\begin{align}
\Psi\left(\dulr,\dulR,t\right) = \sum_{l=1}^{N_{st}}\left|F_l\left(\dulR,t\right)\right|\,\exp{\left[\frac{i}{\hbar}s_l\left(\dulR,t\right)\right]}\varphi_\dulR^{(l)}(\dulr),
\end{align}
the phases of the coefficients in Eq.~(\ref{eqn: born-huang}) and of the coefficients $F_l\left(\dulR,t\right)$ are related via the expression
\begin{align}\label{eqn: phases}
\gamma_l^{(I)}(t) = s_l^{(I)}(t) - S^{(I)}(t).
\end{align}
Here, once again, we have replaced the spatial dependence by the dependence on the trajectory $I$ and we have used the symbol $S^{(I)}(t)$ to indicate the phase of the nuclear wave function, as done in the Letter. Eq.~(\ref{eqn: phases}) straightforwardly follows from the factorization. If we identify the nuclear momentum calculated along the $I$-th trajectory as $\nabla_\nu S^{(I)}(t)+\mathbf A_\nu^{(I)}(t)=\mathbf P_\nu^{(I)}(t)$ and similarly $\nabla_\nu s_l^{(I)}(t)=\mathbf p_\nu^{(l),(I)}(t)$ as the momentum associated to the propagation of the $l$-th ``BO-projected'' wave packet $|F_l^{(I)}(t)|^2$ on the $l$-th BO adiabatic surface, we can write
\begin{align}\label{eqn: gradient of the phases}
\nabla_\nu\gamma_l^{(I)}(t) \simeq \mathbf p_\nu^{(l),(I)}(t) - \left(\mathbf P_\nu^{(I)}(t)-\mathbf A_\nu^{(I)}(t)\right).
\end{align}
Notice that $|F_l^{(I)}(t)|^2$ is identified~\cite{steps, long_steps, long_steps_mt} as a ``BO-projected'' wave packet since from the factorization it follows that $|\chi|^2=\sum_l|F_l|^2$. Taking the time-derivative of both sides of Eq.~(\ref{eqn: gradient of the phases})
\begin{align}\label{eqn: gradient of gamma dot}
\nabla_\nu\dot\gamma_l^{(I)}(t) \simeq -\nabla_\nu\epsilon_{BO}^{(l),(I)}
\end{align}
since on the right-hand-side we identify the forces associated to the motions of the ``BO-projected'' wave packet and of the nuclear wave function. The time-derivative of the term in parenthesis in Eq.~(\ref{eqn: gradient of the phases}) is identically zero because in the chosen gauge the scalar potential is zero, thus all effects of the coupling to the electrons in the nuclear equation is contained in the vector potential. Indeed, Eq.~(\ref{eqn: gradient of gamma dot}) is valid only in the gauge we chose to derive the CT-MQC algorithm. The expression for $\nabla_\nu\gamma_l^{(I)}(t)$ used in our algorithm thus follows,
\begin{align}
\mathbf f_{\nu,l}^{(I)}(t)=\nabla_\nu\gamma_l^{(I)}(t) = -\int^t d\tau\nabla_\nu\epsilon_{BO}^{(l),(I)}.
\end{align}

\section{Approximation for the quantum momentum}
\begin{figure}[h!]
 \begin{center}
 \includegraphics[width=.6\textwidth]{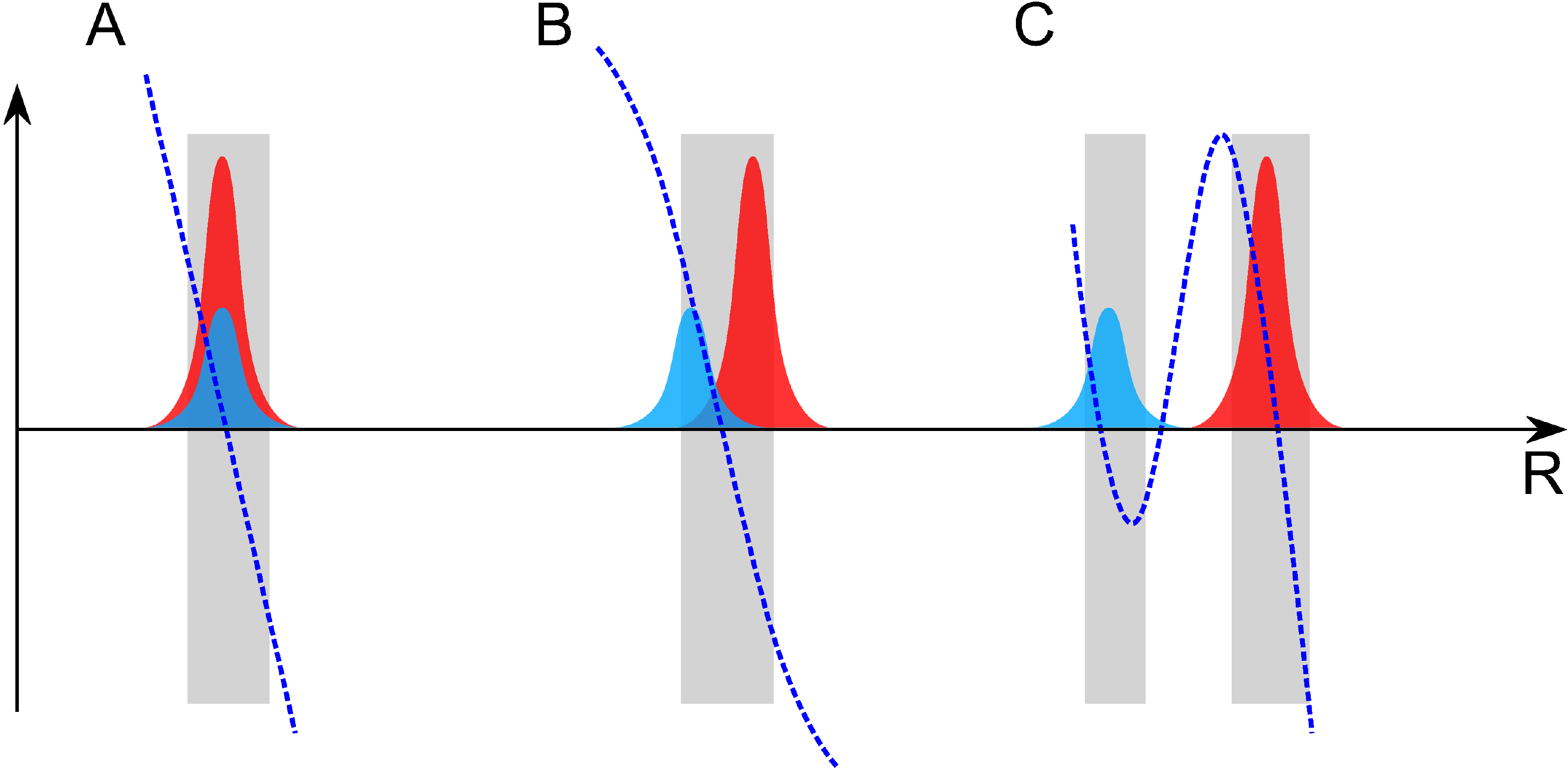}
 \caption{Schematic plot of the shape of the quantum momentum for various cases. The blue and red Gaussians represent BO-projected nuclear densities while the blue dashed lines are the quantum momentum (apart from the $-i\hbar$ multiplicative factor) corresponding to each configuration of the Gaussians. The grey areas highlight the regions where the nuclear density is considerably different from zero.}
 \label{fig: gaussians}
 \end{center}
\end{figure}

We recall here the expression of the quantum momentum
\begin{align}
\boldsymbol{\mathcal P}_\nu^{(I)}(t)=-i\hbar\frac{\nabla_{\nu} \left|\chi^{(I)}(t)\right|}{\left|\chi^{(I)}(t)\right|}=-i\hbar\frac{\nabla_{\nu} \left|\chi^{(I)}(t)\right|^2}{2\left|\chi^{(I)}(t)\right|^2}
\end{align}
which is written in the second equality in terms of the nuclear density $\left|\chi^{(I)}(t)\right|^2$. In order to derive a practical way to estimate this expression, we assume that $\left|\chi^{(I)}(t)\right|^2$ can be written as a sum of Gaussian-shaped wave packets. In the simple case of two adiabatic states, only two Gaussian functions are considered, as schematically illustrated in Fig.~\ref{fig: gaussians}. We underline again (as in the Letter) that this hypothesis is introduced only to estimate the quantum momentum and not to actually solve the classical nuclear equation. We refer to~\cite{semiclassics}, where we show that the approximation we use here for the quantum momentum is consistent also with a representation of the nuclear wave packet in terms of coherent states.
Two scenarios may occur: (1) the BO-projected wave packets overlap and classical trajectories may be located in the overlap region; (2) the BO-projected wave packets are well-separated, thus the overlap region is poorly sampled by the trajectories since the nuclear density is small. In case (1), we use a linear function to approximate the quantum momentum, as shown in the panels (A) and (B) of Fig.~\ref{fig: gaussians}, within the overlap region, while we set it equal to zero outside this region.  Since outside the overlap region there is a low probability of finding trajectories, as the nuclear density is small, the final result is not strongly affected by this approximation. Moreover, as discussed in Eq.~(9) in the Letter, the term in the electronic equation containing the quantum momentum vanishes outside the overlap region, if in this region the non-adiabatic coupling vectors are zero. The inverse situation occurs in case (2), where the linear approximation does not strongly affects the trajectories, because the overlap region is poorly sampled. This is shown in the panel (C) of Fig.~\ref{fig: gaussians}. Outside this region, where the majority of trajectories is located, the quantum momentum is set to zero. However, this situation in general takes place after the nuclear packet has left the region of non-adiabatic coupling, where Eq.~(\ref{eqn: el. additional term}), containing the quantum momentum, is identically zero as discussed in the Letter in Eq.~(9) (this feature is related to the fact that if $|C_l^{(I)}(t)|=1$ all other $|C_k^{(I)}(t)|$ $\forall\,k\neq l$ are identically zero).

\section{Implementing the CT-MQC algorithm}
We describe the steps necessary to implement the CT-MQC:
\begin{enumerate}
\item get the BO potentials $\epsilon_{BO}^{(l),(I)}$, along with $\nabla_\nu\epsilon_{BO}^{(l),(I)}$, and the non-adiabatic coupling vectors $\mathbf d_{\nu,lk}^{(I)}$;
\item compute $\mathbf f_{\nu,l}^{(I)}(t)=\nabla_\nu\gamma_l^{(I)}(t)=-\int^t d\tau\nabla_\nu\epsilon_{BO}^{(l),(I)}$, using the BO forces;
\item collect information about the positions of all trajectories, $\mathbf R^{(I)}(t)$ (this step implies that the trajectories cannot be propagated independently);
\item calculate the quantum momentum, $\nabla_{\nu} \left|\chi^{(I)}(t)\right|/\left|\chi^{(I)}(t)\right| = \alpha\left[R^{(I)}(t) - R_M(t)\right]$, where we use the symbol $R_M(t)$ for the crossing point between the two Gaussians (the $y$-intercept), as illustrated in the Letter, namely by enforcing (the reasonably physical condition) that no population exchange occurs when the non-adiabatic coupling vectors are zero;
\item evolve the coefficients $C_l^{(I)}(t)$ and the trajectories according to Eqs.~(\ref{eqn: el. ehrenfest}) to~(\ref{eqn: n. additional term});
\item go back to step 1.
\end{enumerate}

\section{Computational details}
\subsection{Test case (1)}
\begin{figure}[h!]
 \begin{center}
\includegraphics[width=.7\textwidth]{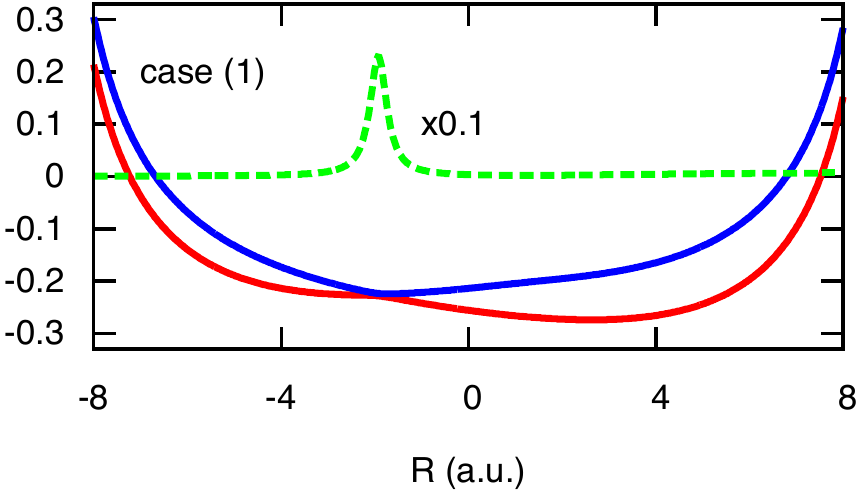}
 \caption{Ground (red) and excited (blue) adiabatic potential energy surfaces for the model case (1). The non-adiabatic coupling between the two states is also shown (green).}
 \label{fig: metiu}
 \end{center}
\end{figure}

The model system for non-adiabatic charge transfer~\cite{MM} analyzed in previous work~\cite{steps, long_steps, long_steps_mt, mqc, *long_mqc, semiclassics} has been also employed in the present study, as it exhibits the typical features of non-adiabatic processes and is exactly solvable, by integrating the full TDSE using the split-operator technique~\cite{spo}. It thus provides a benchmark for the CT-MQC scheme. The system consists of three ions and one electron: two ions are fixed at a distance of $L=19.0$~$a_0$, the third ion and the electron are free to move in one dimension along the line joining the two fixed ions. The moving ion interacts with the fixed ions via a bare Coulomb potentials, while the electron-ion interactions are treated as soft-Coulomb potentials, with parameters $R_m=5.0$~$a_0$ (moving ion), $R_l=3.1$~$a_0$ (left ion) and $R_r=4.0$~$a_0$ (right ion). The full Hamiltonian for this system is
\begin{align}\label{eqn: metiu-hamiltonian}
 \hat{H}(r,R)= &-\frac{1}{2}\frac{\partial^2}{\partial r^2}-\frac{1}{2M}\frac{\partial^2}{\partial R^2}  +
 \frac{1}{\left|\frac{L}{2}
 -R\right|}+\frac{1}{\left|\frac{L}{2} + R\right|}-\frac{\mathrm{erf}\left(\frac{\left|R-r\right|}{R_f}\right)}
 {\left|R - r\right|}\nonumber\\
 &-\frac{\mathrm{erf}\left(\frac{\left|r-\frac{L}{2}
 \right|}{R_r}\right)}{\left|r-\frac{L}{2}\right|}-\frac{\mathrm{erf}\left(\frac{\left|r+\frac{L}{2}\right|}
 {R_l}\right)}{\left|r+\frac{L}{2}\right|}.
\end{align}
The nuclear mass is $M=1836$. With this choice of parameters the two lowest BO states are strongly coupled and there is a weak coupling to the remaining states, as shown in Fig.~\ref{fig: metiu}. The initial wave function is the product of a real-valued normalized Gaussian wave packet, centered at $R_c=-4.0$~$a_0$ with variance $\sigma=1/\sqrt{2.85}$~$a_0$, and the second BO electronic state. The time-step used for integrating the TDSE is $2.4\times10^{-3}$~fs (or $0.1$~a.u.). CT-MQC results are obtained by propagating $N_{traj}=200$ trajectories whose initial positions and momenta are sampled from the Wigner distribution corresponding to the initial quantum mechanical density. In this case, the time-step is $1.2\times10^{-2}$~fs (or $0.5$~a.u.). The forth-order Runge-Kutta algorithm is used to propagate the electronic equation and the velocity-Verlet algorithm is used for the nuclear equation.

\subsection{Test case (2)}
\begin{figure}[h!]
 \begin{center}
\includegraphics[width=.4\textwidth,angle=270]{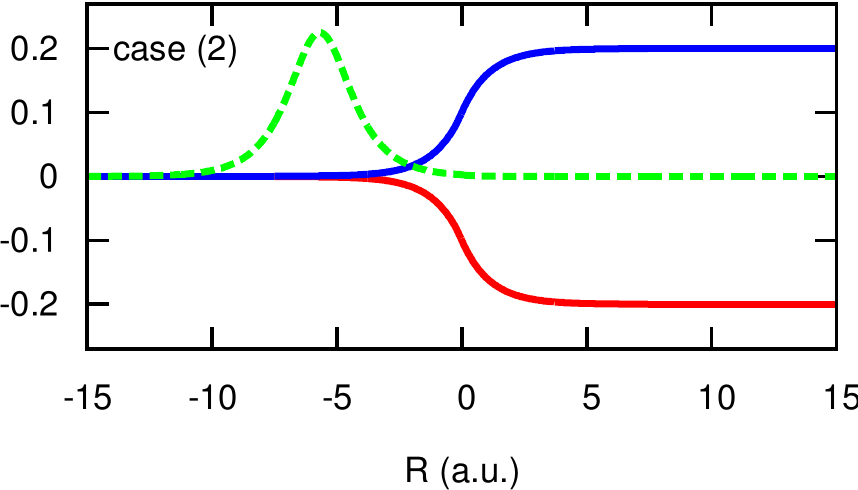}
 \caption{Same as in Fig.~\ref{fig: metiu} but the model case (2).}
 \label{fig: tully 3}
 \end{center}
\end{figure}

In the Letter we have reported results for the so-called Tully's model \#3~\cite{tully1990}. It is a two-state model representing an extended coupling region with reflection. This model has been well-studied in the literature as it represents a critical test for any new approach to non-adiabatic dynamics based on trajectories. In fact, when a nuclear wave packet starts in far negative region and moves with positive but low initial momentum, it is partially reflected by the upper surface and partially transmitted on the lower surface. This event takes place when the nuclear wave packet reaches the region of space where the two adiabatic surfaces start to significantly deviate from each other. The adiabatic potential energy surfaces are shown in Fig.~\ref{fig: tully 3}, along with the non-adiabatic coupling. The analytic form~\cite{tully1990} of the electronic Hamiltonian, in diabatic base, is
\begin{equation}
\begin{array}{lll}
H_{11}(R) &=& a, H_{22}(R) = -a,\\
H_{12}(R) &=& b\exp{\left(cR\right)}, R<0,\\
H_{12}(R) &=& b\left[2-\exp{\left(-cR\right)}\right], R>0,\\
H_{21}(R) &=& H_{12}(R) 
\end{array}
\end{equation}
with $a=6\times 10^{-4},\,b=0.1,\,c=0.9$. In the Letter we show results for a low initial momentum $\hbar k_0=10$~a.u., which represents a critical situation due to the high probability of the reflection channel. The initial nuclear wave packet is as Gaussian with width $\sigma=20/k_0$ with $R_0=-15.0$~a.u. the center of the Gaussian. Exact results are obtained using a wave packet propagation scheme where we chose the value 0.1~a.u. for the time-step used to integrate the TDSE. CT-MQC results are obtained by propagating a set of $N_{traj}=200$ trajectories whose initial positions and momenta are sampled from the Wigner distribution corresponding to the initial quantum mechanical density. The time-step is $0.5$~a.u. for the CT-MQC algorithm.

\subsection{Analysis in momentum space}
\begin{figure}[h!]
 \begin{center}
\includegraphics[width=.6\textwidth]{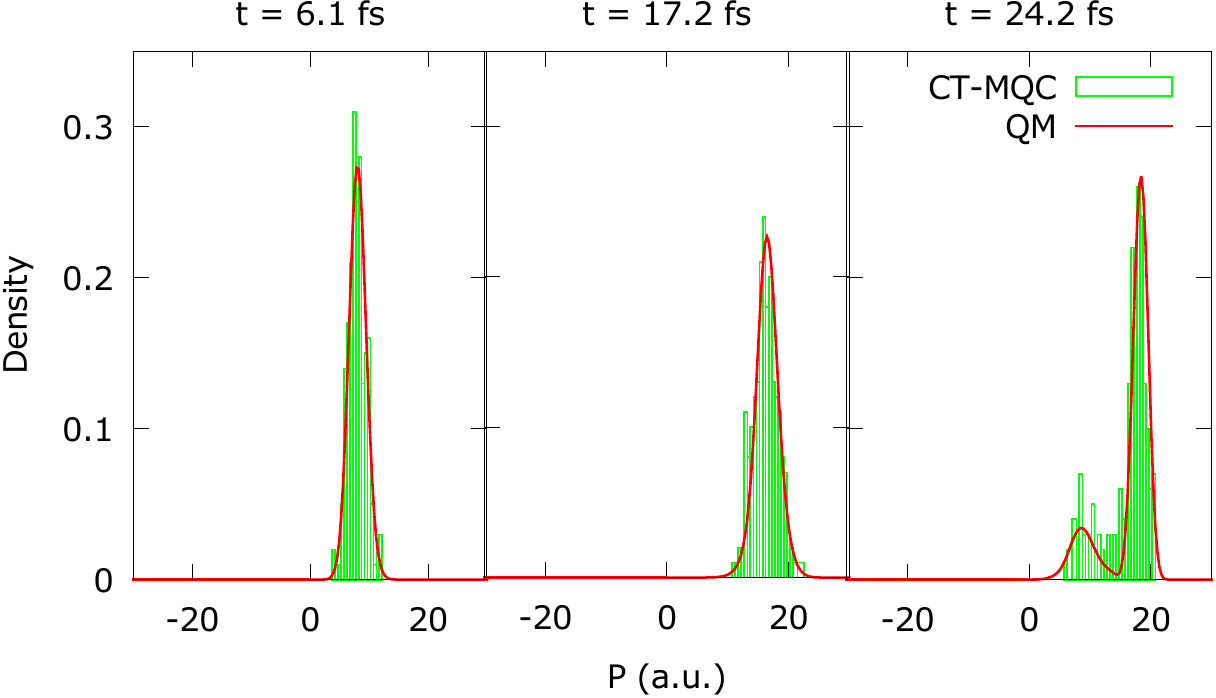}
 \caption{Splitting of the nuclear density in momentum representation (line), compared with quantum-classical results (histogram).}
 \label{fig: momentum splitting}
 \end{center}
\end{figure}

We have observed in the Letter that decoherence starts manifesting itself when the nuclear density splits in configuration space. The two events, in fact, are strongly related: decoherence cannot be appropriately reproduced if the wave packets propagates ``coherently'', without splitting. However, the deficiency of TSH in reproducing decoherence proves that we can observe splitting without decoherence. Despite this last observation we find interesting to look at the splitting of the nuclear density not only in configuration space, but also in momentum space. The position representation seems indeed more natural, due to the fact the BO states are only defined ``at fixed nuclear positions'' and not ``at fixed nuclear momenta''. In Fig.~\ref{fig: momentum splitting} we show the splitting of the nuclear density in momentum space, which is perfectly captured by the momentum distribution of the classical trajectories.

\subsection{Computational cost of the CT-MQC scheme}
The actual bottleneck of trajectory-based approaches to non-adiabatic dynamics is the computation of electronic structure properties. However, the solution of the full quantum problem requires to treat the nuclei quantum-mechanically and indeed here a trajectory-based method is able to cut the computational cost.

For instance, we have compared in the proton-coupled-electron-transfer problem the computational cost needed to solve the TDSE with the CT-MQC equations. The TDSE is integrated using split-operator-technique, referenced in the Supplemental Material, while the electronic and nuclear equations of the CT-MQC scheme are solved with the fourth-order Runge-Kutta algorithm and velocity-Verlet algorithm, respectively. In these particular cases, quantum mechanical calculations require a smaller integration time-step than the quantum-classical method, i.e. 5 times smaller for the case shown in the Letter. Solving the TDSE up to 1000~a.u. with $1000\times1000$ grid points for one nucleus and one electron takes approximately 15 minutes on a normal desktop (Intel Core i7-4790 3.60 GHz), while it takes 1 minute for the CT-MQC with 200 trajectories. To perform the same number of integration steps, the computational cost of the CT-MQC is approximately a factor 3 smaller than solving the TDSE.

As for the scaling of the computational cost in going to higher dimensions, the numerical effort needed to include a larger number of trajectories, in order to appropriately sample the configurations space, scales linearly with the number of trajectories.

\section{Time-dependent vector potential}
\begin{figure}[h!]
 \begin{center}
\includegraphics[width=.6\textwidth]{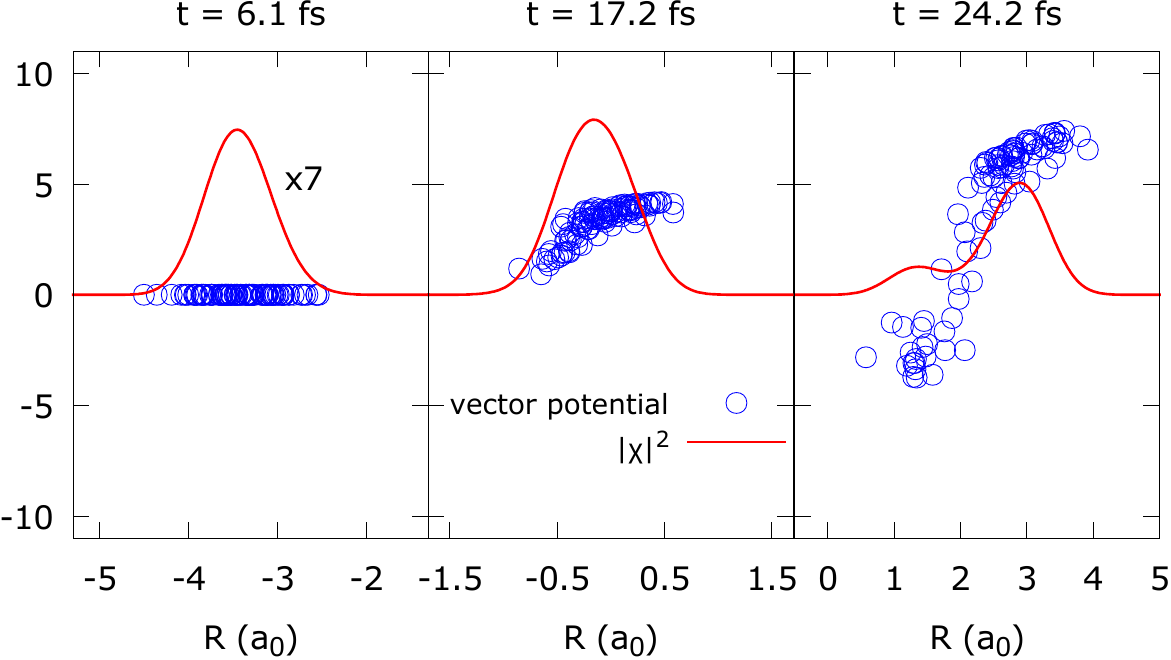}
 \caption{Snapshots of the vector potential taken along the quantum-classical evolution.}
 \label{fig: vector potential}
 \end{center}
\end{figure}

The time-dependent vector potential of the theory is a gauge-dependent quantity. Exact quantum-mechanical results are presented in a gauge where the vector potential is set to zero (that is possible since the systems under considerations are solved in one-dimension). In order to propose a CT-MQC procedure as general as possible, we have chosen to work in a different gauge when developing the algorithm, namely
\begin{align}
\epsilon_0^{(I)}(t)+\epsilon_{\mathrm{TD}}^{(I)}(t)+\sum_\nu\mathbf V_\nu^{(I)}\cdot\bA_\nu^{(I)}(t)=0
\end{align}
Therefore, as an example, we show in Fig.~\ref{fig: vector potential} the vector potential calculated within the CT-MQC scheme for model case (1). It cannot be compared with exact results, as it is identically zero in the quantum case. This is also the reason why the only potential shown in the Letter is the gauge-invariant part of the TDPES.
The vector potential is non-zero only after the nuclear density splits in two branches. Notice that it appears as a nuclear momentum correction, due to the non-adiabatic nature of the electrons, in the nuclear Hamiltonian in Eq.~(2) of the Letter. At time $t=24.2$~fs in Fig.~\ref{fig: vector potential}, the vector potential contributes a negative term only to those trajectories that move on the upper BO surface, as its negative part corresponds to the region occupied by the BO-projected wave packet ``on'' the upper BO surface. The opposite situation applies to the positive part of the vector potential. This observation is in agreement with previous analysis reported in Ref.~\cite{long_steps_mt}.

\addcontentsline{toc}{section}{References}
%